\documentclass{cimento}
\usepackage{graphicx}

\def\as{\alpha_S}
\def\eps{\varepsilon}
\usepackage{cite}

\title{Further exploration of top pair hadroproduction at NNLO}

\author{
M.~Czakon\from{ins:Aachen},
P.~Fiedler\from{ins:Aachen},
A.~Mitov\from{ins:CERN}\thanks{Speaker} and 
J.~Rojo\from{ins:CERN}
}

\instlist{
\inst{ins:Aachen} Institut f\"ur Theoretische Teilchenphysik und Kosmologie, RWTH Aachen University\\ D-52056 Aachen, Germany
\inst{ins:CERN}Theory Division, CERN, CH-1211 Geneva 23, Switzerland
}

\PACSes{PACS: 14.65.Ha, 13.85.Lg, 12.38.Bx\\ Preprint: CERN-PH-TH/2013-100,  TTK-13-15}

\begin{document}

\maketitle

\begin{abstract}
Top quark pair production is one of the cornerstones of the physics program at
hadron colliders. In this contribution, we further explore the phenomenological  implications of
the recent NNLO calculation of the total inclusive cross-section. We provide
a comparison of  the scale dependence of the top pair hadroproduction cross section at different perturbative orders and study its perturbative convergence (with and without soft-gluon resummation). We also sketch how the NNLO top quark cross section could be used to improve searches of physics beyond the Standard Model.
\end{abstract}

\section{Introduction}

Top quark pair production is one of the cornerstones of the Standard Model (SM) program at hadron colliders, and a number of precision calculations of this process have appeared in the recent past. In this writeup, we focus our attention on the total inclusive cross-section which, during the last year, became known in full NNLO \cite{Baernreuther:2012ws,Czakon:2012zr,Czakon:2012pz,Czakon:2013goa},
and present analyses based on the NNLO calculation that are not available in the literature. 
\footnote{For a broader recent overview of theoretical developments in top quark physics see, for example, Ref.~\cite{Mitov:2013jla}.}

This writeup is organized as follows: in section \ref{sec:notations} we introduce our notation. In section \ref{sec:factorization} we give the explicit results for the collinear factorization contribution and for the scale dependent terms in the $gg$ reaction, both of which were not explicitly presented in Ref.~\cite{Czakon:2013goa}. In section \ref{sec:perturbative} we present a number of results that illustrate the convergence properties of perturbation theory with and without soft-gluon resummation. Finally in section \ref{sec:stop} we present some preliminary results that illustrate how precision top pair production can be relevant for searches of BSM physics.

\section{The $t\bar t$ total cross-section: notations}\label{sec:notations}

We follow the notation established in Refs.~\cite{Baernreuther:2012ws,Czakon:2012zr,Czakon:2012pz,Czakon:2013goa}. The total inclusive top pair production cross-section is defined as
\begin{equation}
\sigma_{\rm tot} = \sum_{i,j} \int_0^{\beta_{\rm max}}d\beta\, \Phi_{ij}(\beta,\mu_F^2)\, \hat\sigma_{ij}(\as(\mu_R^2),\beta,m^2,\mu_F^2,\mu_R^2)\, .
\label{eq:sigmatot}
\end{equation}
The indices $i,j$ run over all possible initial state partons; $\beta_{\rm max} \equiv \sqrt{1-4m^2/S}$; $\sqrt{S}$ is the c.m. energy of the hadron collider and $\beta=\sqrt{1-\rho}$, with $\rho\equiv 4m^2/s$, is the relative velocity of the final state top quarks with pole mass $m$ and partonic c.m. energy $\sqrt{s}$. 

The function $\Phi$ in Eq.~(\ref{eq:sigmatot}) is the partonic flux 
\begin{equation}
\Phi_{ij}(\beta,\mu_F^2) = {2\beta \over 1-\beta^2}~ {\cal L}_{ij}\left({1-\beta_{\rm max}^2\over 1-\beta^2}, \mu_F^2\right) \, ,
\label{eq:flux}
\end{equation}
expressed through the partonic luminosity
\begin{equation}
{\cal L}_{ij}(x,\mu_F^2) = x \left( f_i\otimes f_j \right) (x,\mu_F^2) = x \int_0^1 dy \int_0^1 dz \, \delta(x-yz) f_i(y,\mu_F^2)f_j(z,\mu_F^2) \, .
\label{eq:Luminosity}
\end{equation}

As usual, $\mu_{R,F}$ are the renormalization and factorization scales. Setting $\mu_F=\mu_R=\mu$, the partonic cross-section can be expanded through NNLO as
\begin{eqnarray}
\hat\sigma_{ij} = {\as^2\over m^2}\Bigg\{  \sigma^{(0)}_{ij} + \as \left[ \sigma^{(1)}_{ij} + L\, \sigma^{(1,1)}_{ij} \right] + \as^2\left[ \sigma^{(2)}_{ij} + L\, \sigma^{(2,1)}_{ij} + L^2 \sigma^{(2,2)}_{ij} \right] \Bigg\} \, .
\label{eq:sigmapart}
\end{eqnarray}
In the above equation $L = \ln\left(\mu^2/m^2\right)$, $\as$ is the ${\overline {\rm MS}}$ coupling renormalized with $N_L=5$ active flavors at scale $\mu^2$ and $\sigma^{(n(,m))}_{ij}$ are functions only of $\beta$.

All partonic cross-sections are known through NNLO \cite{Baernreuther:2012ws,Czakon:2012zr,Czakon:2012pz,Czakon:2013goa}. The scaling functions $\sigma^{(2,1)}_{ij}$ and $\sigma^{(2,2)}_{ij}$ can be computed from $\sigma^{(1)}_{ij}$, see section \ref{sec:factorization}. The dependence on $\mu_R\neq \mu_F$ can be trivially restored in Eq.~(\ref{eq:sigmapart}) by re-expressing $\as(\mu_F)$ in powers of $\as(\mu_R)$; see for example Ref.~\cite{Langenfeld:2009wd}.

\section{Collinear factorization and scale dependence of the partonic cross-section}\label{sec:factorization}\label{sec:factorization}

We follow the setup and notation described in Ref.~\cite{Czakon:2012zr} and denote the collinearly unrenormalized partonic cross-sections as $\tilde \sigma_{ij}^{(n)}(\eps,\rho)$. Then, introducing the functions $\tilde s^{(n)}_{ij}$ and $s^{(n)}_{ij}$ defined as $\tilde s_{ij}^{(n)}(\eps,\rho) \equiv \tilde\sigma^{(n)}_{ij}(\eps,\rho)/\rho$ and $s^{(n)}_{ij}(\rho) \equiv \sigma^{(n)}_{ij}(\rho)/\rho$, the $\overline{\rm MS}$--subtracted $gg$-initiated cross-section $s^{(n)}_{gg}$ reads through NNLO:
\begin{eqnarray}
s^{(1)}_{gg} &=& \tilde s^{(1)}_{gg} + {2\over \epsilon} \left({1\over 2\pi}\right) \tilde s^{(0)}_{gg} \otimes P^{(0)}_{gg} \, , \label{eq:shat1}\\ 
&& \nonumber\\
s^{(2)}_{gg} &=& \tilde s^{(2)}_{gg} +  \left({1\over 2\pi}\right)^2 \Bigg\{ {1\over \eps^2}\left[ -\beta_0\tilde s^{(0)}_{gg} \otimes P^{(0)}_{gg} + 2 \tilde s^{(0)}_{gg} \otimes P^{(0)}_{gg}\otimes P^{(0)}_{gg}\right. \label{eq:shat2}\\
&&\left. + 2N_L\left(\tilde s^{(0)}_{gg} \otimes P^{(0)}_{gq}\otimes P^{(0)}_{qg} + \tilde s^{(0)}_{q \bar q} \otimes P^{(0)}_{qg}\otimes P^{(0)}_{qg}\right) \right] + {1\over \epsilon}\tilde s^{(0)}_{gg} \otimes P^{(1)}_{gg}\Bigg\}\nonumber\\
&&  +  {1\over \epsilon} \left({1\over 2\pi}\right) \Bigg\{ 4N_L \tilde s^{(1)}_{qg} \otimes P^{(0)}_{qg} + 2\tilde s^{(1)}_{gg} \otimes P^{(0)}_{gg}\Bigg\} \, ,\nonumber
\end{eqnarray}
with $\beta_0=11C_A/6-N_L/3$.

The integral convolutions in Eq.~(\ref{eq:shat2}) are performed numerically, over a set of 80 points in the interval $\beta \in (0,1)$.  The calculation of the partonic cross-section $\tilde s^{(1)}_{gg}$ through order ${\cal O}(\epsilon)$ has been detailed in Ref.~\cite{Czakon:2012pz}.

The evaluation of the scale dependent functions $\sigma^{(2,1)}_{gg}$ and $\sigma^{(2,2)}_{gg}$ in Eq.~(\ref{eq:sigmapart}) is rather straightforward, see \cite{Czakon:2012zr} for details. In terms of the functions $s^{(n(,m))}_{ij}(\rho) \equiv \sigma^{(n(,m))}_{ij}(\rho)/\rho$ we get:
\begin{eqnarray}
s^{(2,2)}_{gg} &=& {1\over (2\pi)^2} \Bigg\{
3\beta_0^2 s^{(0)}_{gg} -5 \beta_0s^{(0)}_{gg}\otimes P^{(0)}_{gg} + 2 s^{(0)}_{gg}\otimes P^{(0)}_{gg}\otimes P^{(0)}_{gg}\nonumber\\
&&+2N_L\left(  s^{(0)}_{gg}\otimes P^{(0)}_{qg}\otimes P^{(0)}_{gq}+s^{(0)}_{q\bar q}\otimes P^{(0)}_{qg}\otimes P^{(0)}_{qg}\right)\Bigg\} \, ,\nonumber\\
s^{(2,1)}_{gg} &=& {2\over (2\pi)^2} \Bigg\{ \beta_1 s^{(0)}_{gg} - s^{(0)}_{gg}\otimes P^{(1)}_{gg}\Bigg\}\nonumber\\
&&+ {1\over 2\pi} \Bigg\{ 3\beta_0 s^{(1)}_{gg} - 2s^{(1)}_{gg}\otimes P^{(0)}_{gg} - 4N_L s^{(1)}_{qg}\otimes P^{(0)}_{qg}\Bigg\}\, ,
\label{scales-qtildeq}
\end{eqnarray}
with $\beta_1 = 17 C_A^2/6 - 5 C_A N_L/6 - C_F N_L/2$.

Eq.~(\ref{scales-qtildeq}) agrees with Ref.~\cite{Langenfeld:2009wd}. The convolutions appearing in Eq.~(\ref{scales-qtildeq}) are computed numerically. High quality fits to the functions $\sigma^{(2,1)}_{gg}$ and $\sigma^{(2,2)}_{gg}$ have been implemented in version 2.0 of the program {\tt Top++} \cite{Czakon:2011xx}
\footnote{Fits implemented in the program Hathor \cite{Aliev:2010zk} have also been utilized.}
and can be read off from there.

\section{Perturbative convergence of the hadronic cross-section}\label{sec:perturbative}

The size of the scale dependence of the $t\bar t$ cross-section at NNLO and NNLO+NNLL has been studied in Ref.~\cite{Czakon:2013goa},  while a  detailed breakdown of the various sources of theoretical uncertainty (PDFs, scale,
$\alpha_s$ and $m_{\rm top}$) was provided in Ref.~\cite{Czakon:2013tha}. In the following we will study the changes of the scale dependence of the total cross-section as a function of the perturbative order. As a representative case, we focus our discussion on LHC 8 TeV. We also update the corresponding plot for the Tevatron from Ref.~\cite{Baernreuther:2012ws}. 

We begin by first comparing the pure fixed order predictions i.e. not including soft gluon resummation. We compare the LO, NLO and NNLO results, and each one is computed with a PDF set of matching accuracy. For consistency with our earlier presentations we use everywhere the MSTW2008 (68cl) family of PDF sets~\cite{Martin:2009iq}. Similar results are obtained if other PDF sets such as CT10~\cite{Gao:2013xoa}  and NNPDF2.3~\cite{Ball:2012cx}  are used, see Ref.~\cite{Czakon:2013tha} for a detailed comparison of the predictions from the various sets.

\begin{figure}
\begin{center}
\includegraphics[width=6.8cm]{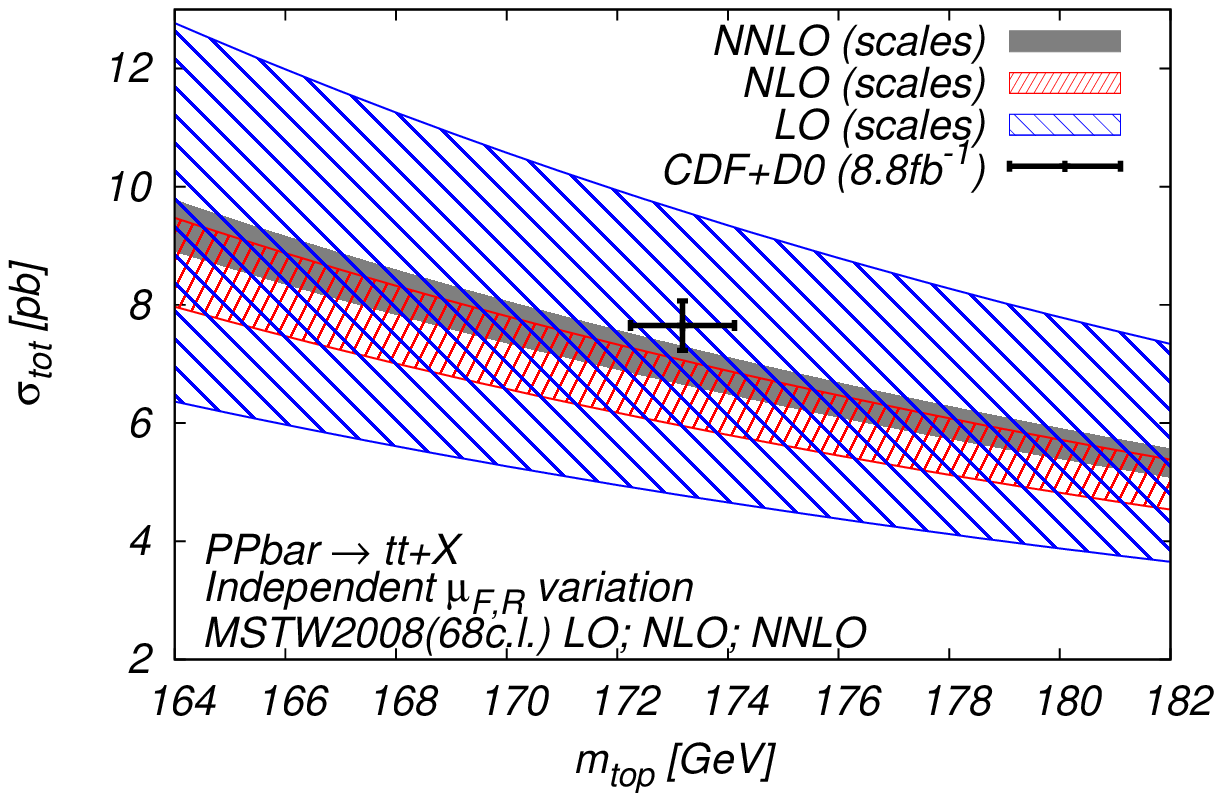}
   \hspace{-.4cm} 
\includegraphics[width=6.8cm]{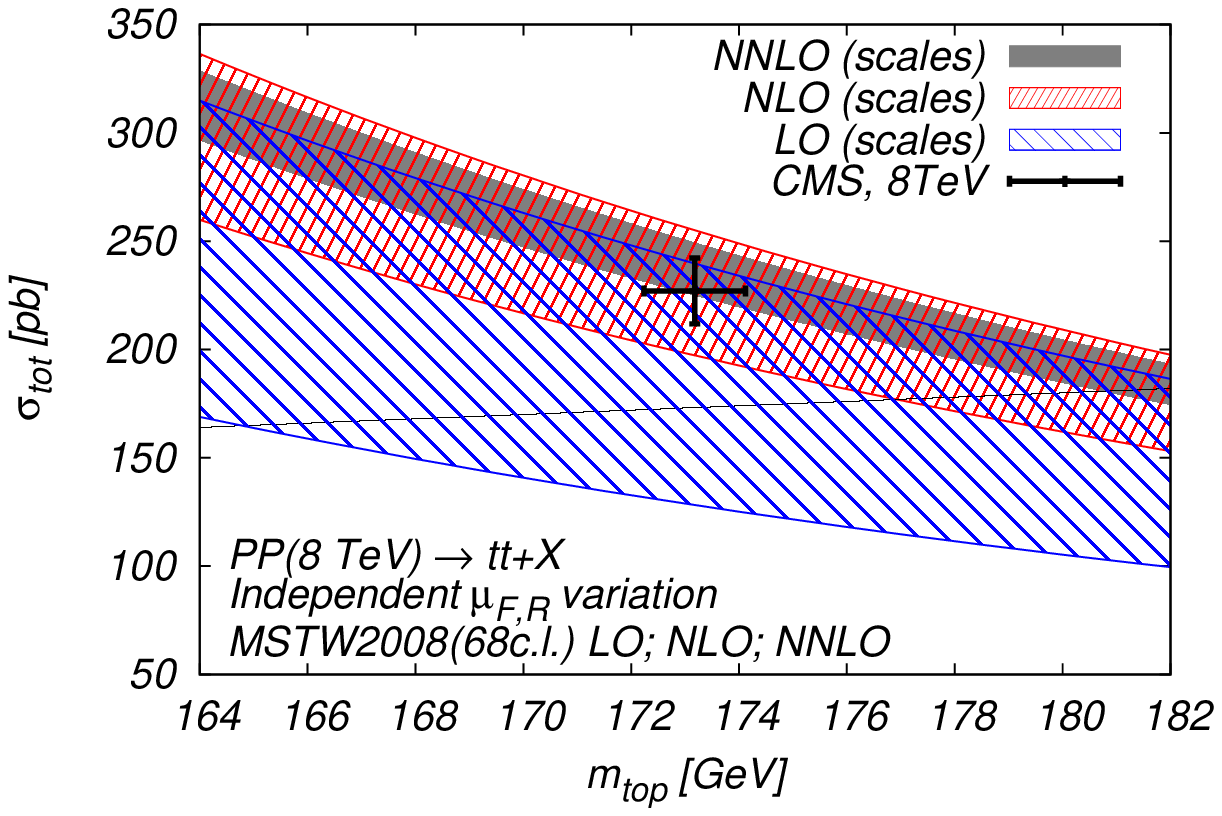}
\caption{Scale dependence of the total cross-section at LO (blue), NLO (red) and NNLO (black) as a function of $m_{\rm top}$ at the Tevatron (left) and the LHC 8 TeV (right). No soft gluon resummation is included. For reference the most precise experimental measurements are also shown.}
\label{fig:tev+lhc-mass}
\end{center}
\end{figure}

In fig.~\ref{fig:tev+lhc-mass} (left) we show the scale dependence of the predicted cross-section at the Tevatron, as a function of the top quark mass. We note the significant and consistent improvement in the theoretical precision due to inclusion of corrections at higher perturbative orders. We also note the agreement between the theoretical prediction 
\footnote{Recall that only the scale dependence is shown. The full theoretical uncertainty is, roughly, about twice as large as the scale dependence.}
and the latest Tevatron measurement \cite{tev-sigma_exp}. 

Next we turn to the LHC. In fig.~\ref{fig:tev+lhc-mass} (right) we show the scale dependence of the predicted cross-section at the LHC 8 TeV as a function of $m_{\rm top}$.  Similarly to the case of the Tevatron, we observe a very good perturbative convergence of the theoretical prediction and good agreement with the available measurement \cite{CMS8TeV}. 

\begin{figure}
\begin{center}
\includegraphics[width=6.8cm]{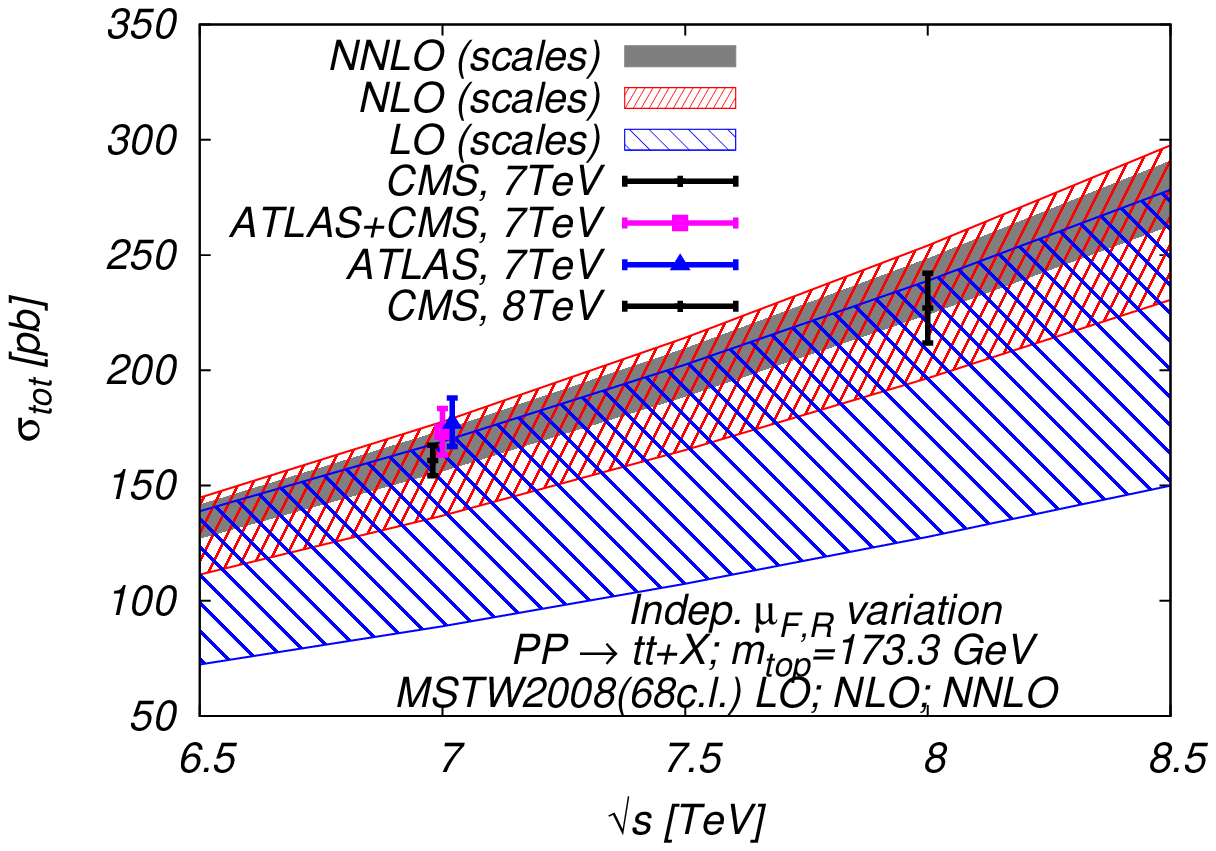}
   \hspace{-.4cm}
\includegraphics[width=6.8cm]{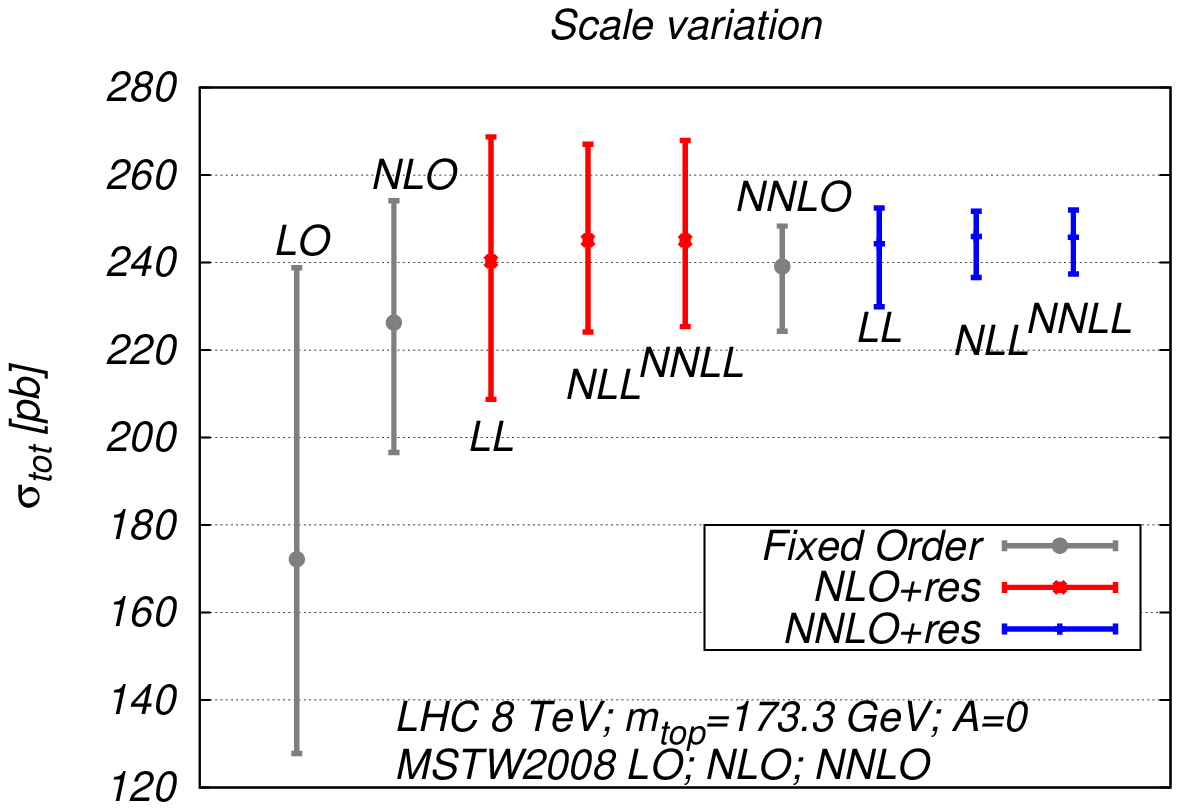}
\caption{Scale dependence of the predicted cross-section at LO, NLO and NNLO at the LHC as a function of $\sqrt{s}$ (left).  On the right plot: detailed breakdown of scale uncertainty for LHC 8 TeV at LO, NLO and NNLO including also soft-gluon resummation at LL, NLL and NNLL.}
\label{fig:tev+lhc}
\end{center}
\end{figure}

In fig.~\ref{fig:tev+lhc} (left) we show the scale dependence of the predicted cross-section at the LHC as a function of the collider energy. We note that the perturbative convergence observed at 8 TeV is consistently present in the whole range of relevant LHC energies. Moreover, the good agreement of the NNLO theoretical prediction with the available data persists at all energies where data is currently available \cite{Atlas7TeV,Chatrchyan:2012bra,Atlas_and_CMS-7TeV}.

Next we study the impact of soft-gluon resummation on the size of the scale dependence and the central value of the theoretical prediction. In fig.~\ref{fig:tev+lhc} (right) we show the scale dependence of the predicted cross-section at the LHC 8 TeV for a number of cases with different fixed order and logarithmic accuracy: LO, NLO, NLO+LL, NLO+NLL, NLO+NNLL, NNLO, NNLO+LL, NNLO+NLL and NNLO+NNLL. In all cases we follow the resummation procedure of Ref.~\cite{Cacciari:2011hy}. We set the constant $A=0$ (introduced in Ref.~\cite{Bonciani:1998vc}), $m_{\rm top}=173.3$ GeV and set the accuracy of the pdf according to the accuracy of the fixed order result. 

We observe that the excellent convergence of the perturbative expansion is preserved after the inclusion of soft gluon resummation. In particular, the feature that resummation shifts the fixed order cross-section up by about 2-3\% is consistently present at NLO and NNLO and does not seem to significantly depend on the logarithmic accuracy of the resummation. Inclusion of resummation with logarithmic accuracy at NLL or NNLL also noticeably decreases the scale dependence of the theoretical prediction, as expected. The absolute size of the resulting reduction in scale dependence is also at the 2\% level. 

\begin{figure}
\begin{center}
\includegraphics[width=6.8cm]{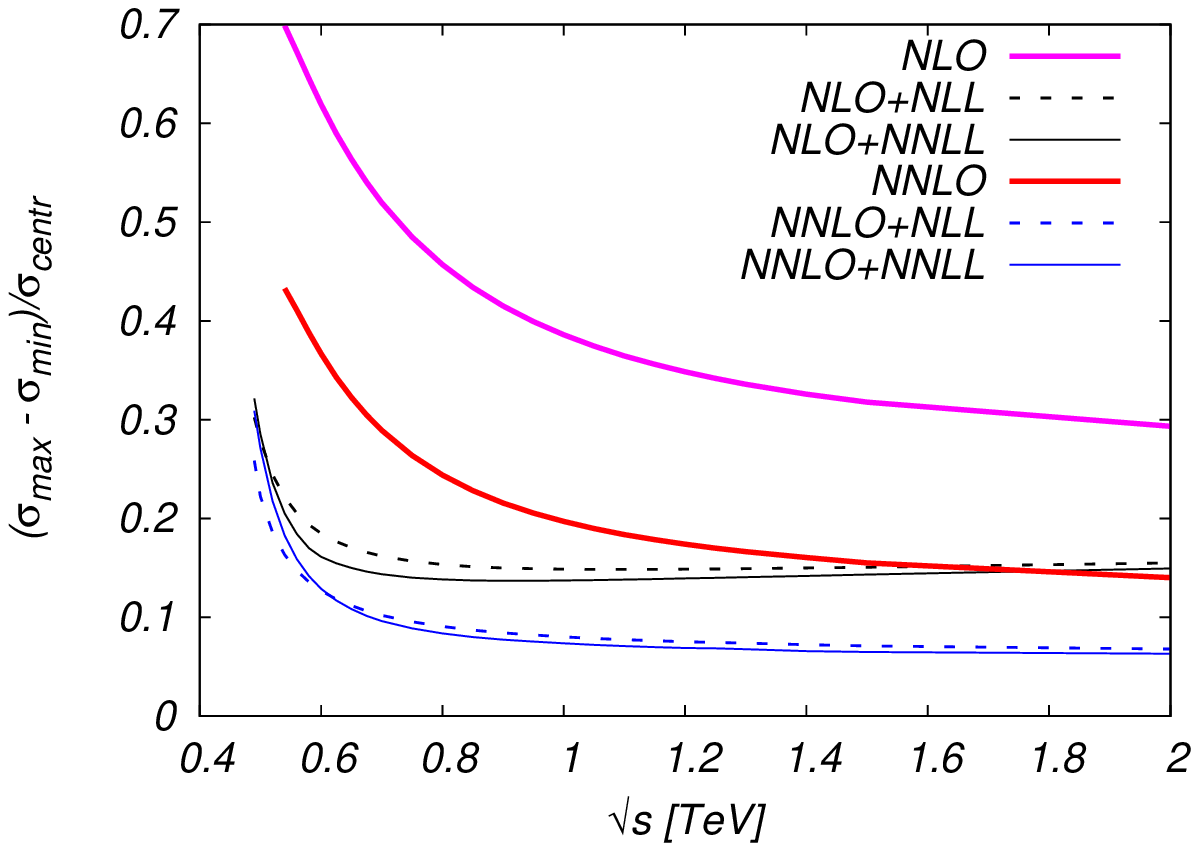}
   \hspace{-.4cm}
\includegraphics[width=6.8cm]{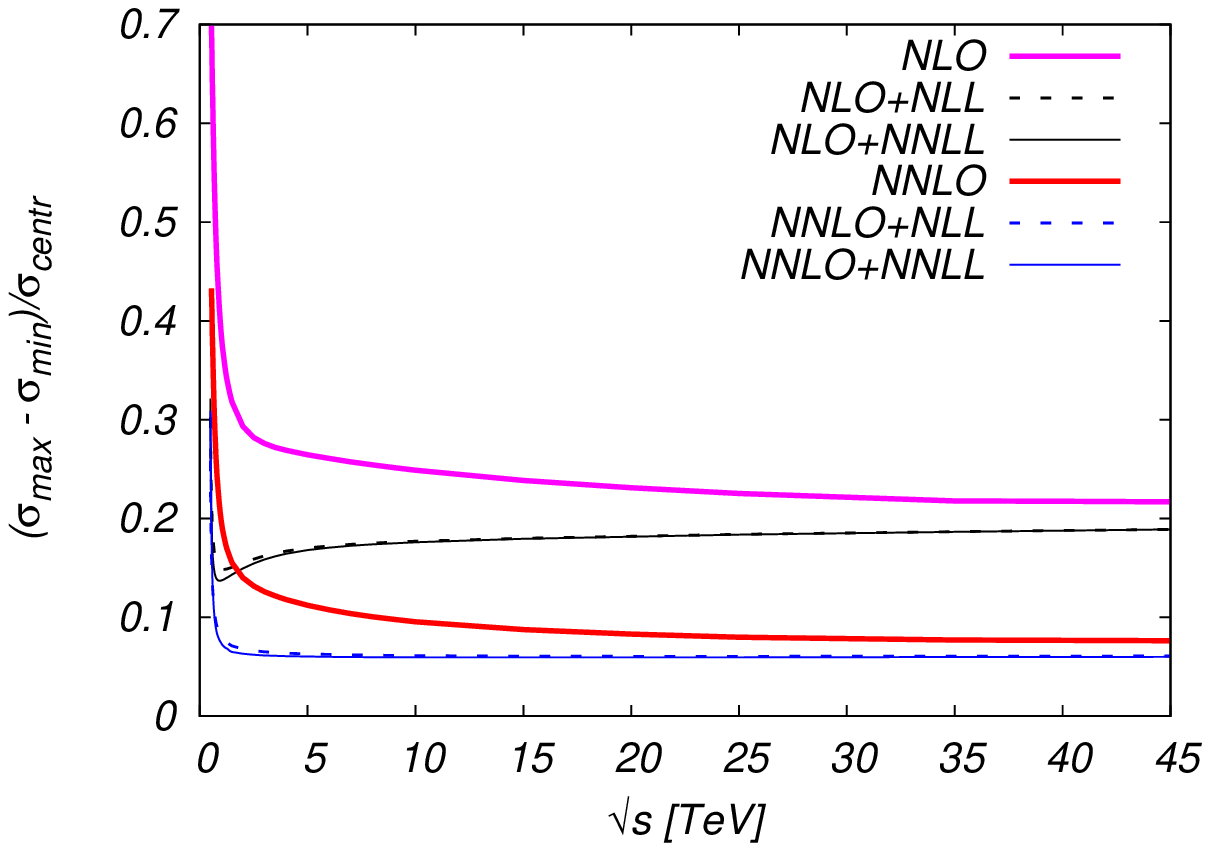}
\caption{The relative scale uncertainty of the $t\bar t$ cross-section, computed as a function of the LHC collider energy at fixed order (NLO and NNLO) and including with soft-gluon resummation (NLL and NNLL).}
\label{fig:compare}
\end{center}
\end{figure}

An alternative way of assessing the impact of soft-gluon resummation is shown in fig.~\ref{fig:compare} (which updates fig.~1 of Ref.~\cite{Cacciari:2011hy} by including the exact NNLO result). Plotted is the relative error of the cross-section at the LHC as a function of the collider energy. We consider a broad range of energies, starting from slightly above the $t\bar t$ production threshold and going up to 45 TeV which is far above threshold. In all cases we observe that the inclusion of soft gluon resummation extends the validity of the perturbative prediction closer to threshold. For large collider energies the enhanced $t\bar t$ threshold contribution gets reduced and, indeed, we observe that the resummed and unresummed predictions converge to each other in this case. We also notice that the difference between NLL and NNLL is small and is more pronounced when added on top of the NLO result (as anticipated). Finally we note that the inclusion of soft-gluon resummation on top of the NNLO result makes the relative scale uncertainty practically independent of the collider energy, except of course for the immediate threshold region which, {\it a posteriori}, is another justification for the use of soft-gluon  resummation.

\section{Application to searches for physics beyond the Standard Model}\label{sec:stop}

In addition to being a powerful tool for testing the Standard Model, the high precision of the total inclusive $t\bar t$ production cross-section presents an opportunity for devising new strategies for searches of physics beyond the Standard Model. 
A first exploration of the improvements in BSM searches arising from NNLO top data was presented in Ref.~\cite{Czakon:2013tha}, where it was shown that the use of top quark data in a NNLO global PDF fit leads to an improved determination of the poorly known large-$x$ gluon PDF. This improvement then translates into more accurate predictions for BSM heavy particle production 
and for the large mass tail of the $M_{tt}$ distribution, the latter used in searches of new heavy resonances which decay into top quarks.

\begin{figure}
  \begin{center}
\includegraphics[width=6.8cm]{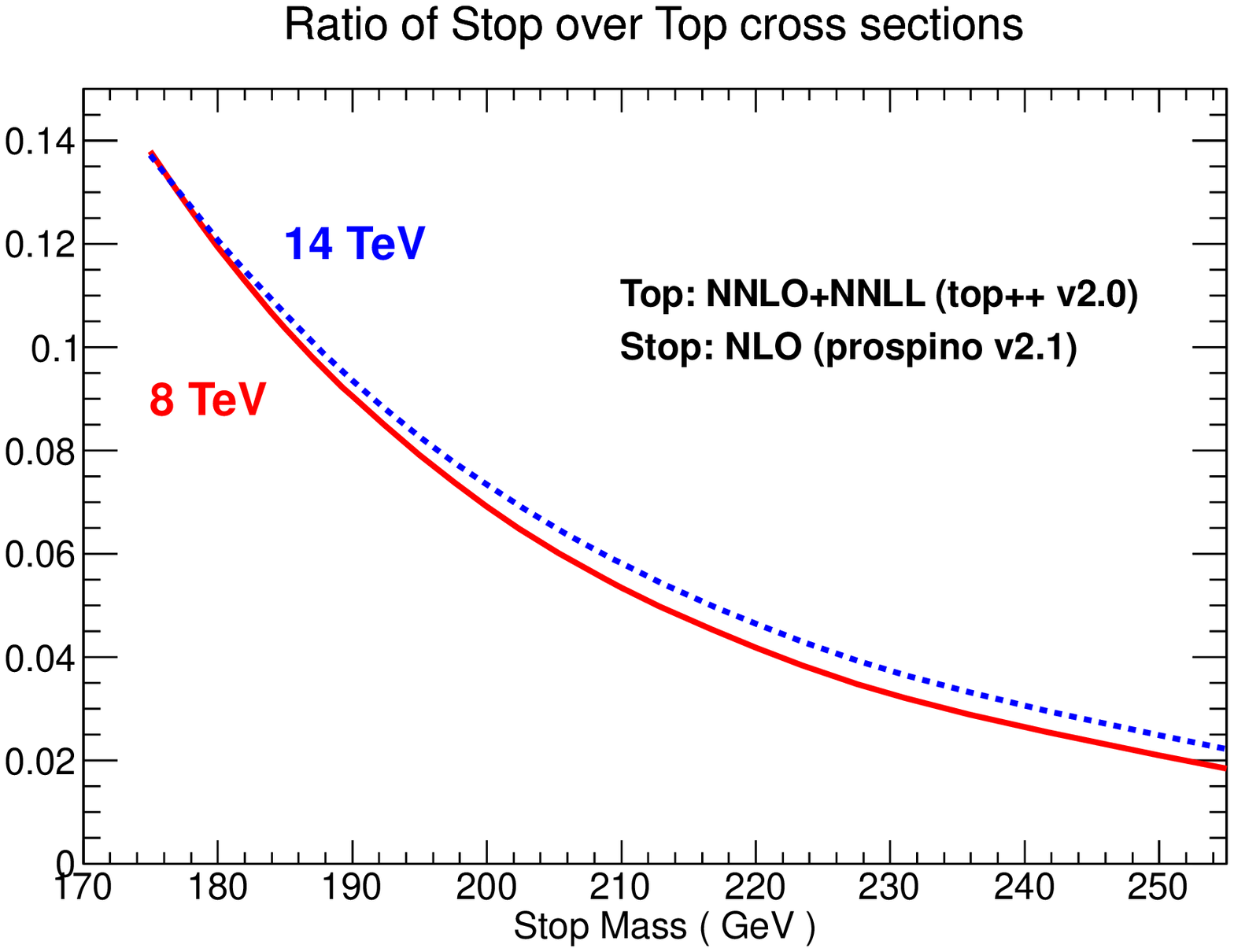}
   \hspace{-.4cm} 
\includegraphics[width=6.8cm]{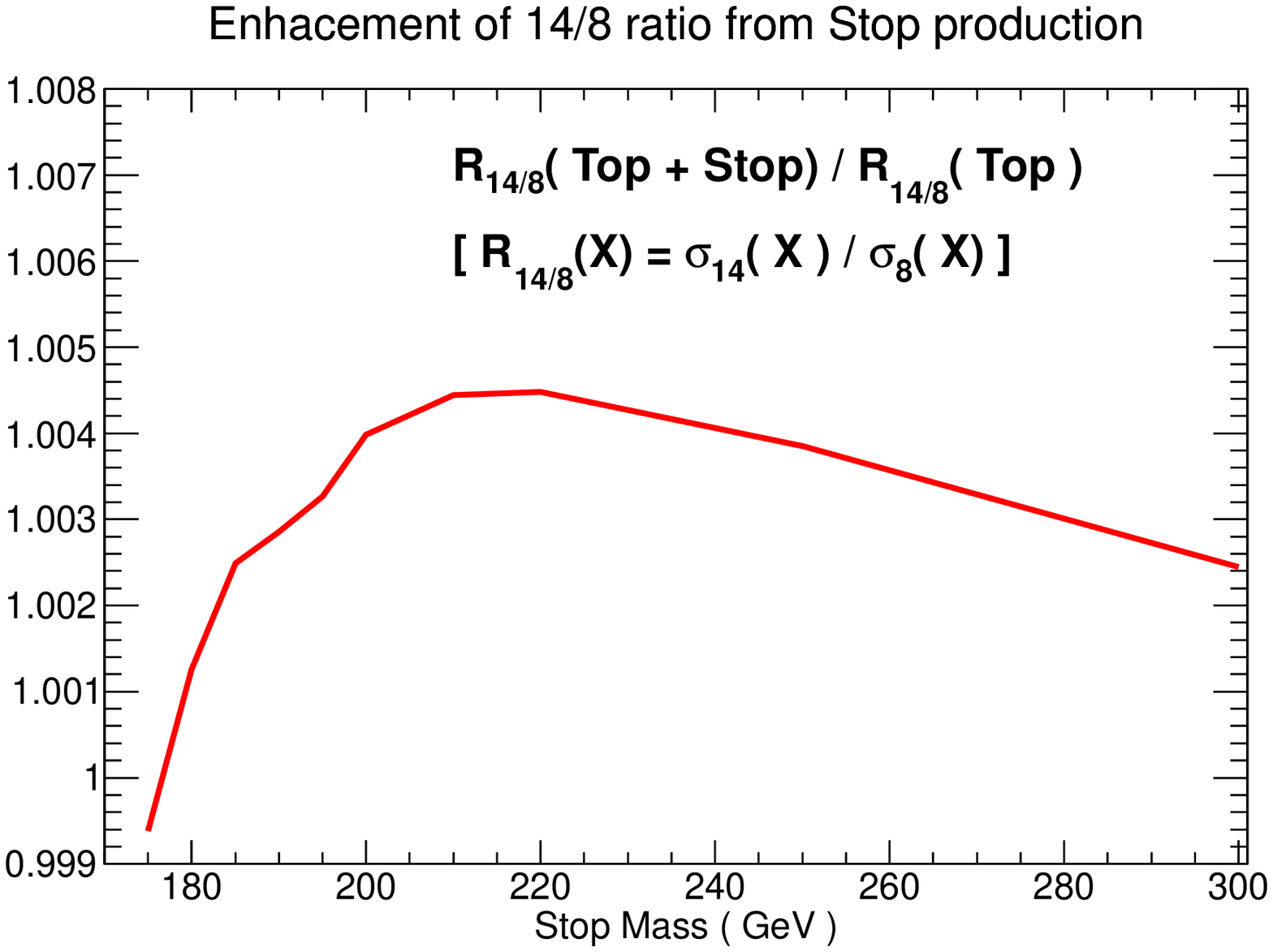}
\caption{Stop production at LHC 8 and 14 TeV. Left plot: the ratio of the stop and top production cross-sections. Right plot: the double ratio of the sum of top and stop cross-sections at 8 and 14 TeV normalized to pure top pair cross-section at 8 and 14 TeV. The top pair cross-section is evaluated at NNLO+NNLL with {\tt Top++(2.0)} while the stop pair cross-section is evaluated at NLO with the help of the program {\tt Prospino(2.1)}.}
\label{fig:stop}
\end{center}
\end{figure}

While the above examples illustrate the indirect improvement in BSM searches due to top quark data,
high-precision top production can also impact BSM studies directly, for example, in the search for supersymmetric top partners - the stops. 
The basic idea is rather simple \cite{CMRPW}: in searches for stops with mass that is only slightly above the top mass, the stops decay to either a pair of top quarks or to the decay products of the top quark. Either way, the conventional stop searches require separation of the stop signal from the very similar and much larger top background. The ratio of the stop over top cross-sections is shown in fig.~\ref{fig:stop} (left) for
LHC 8 and 14 TeV. The computation of the top cross-section is done at NNLO+NNLL with the program {\tt Top++ (2.0)} \cite{Czakon:2011xx}, while the stop cross-section is computed at NLO with the program {\tt Prospino(2.1)} \cite{Beenakker:1997ut},
using consistently MSTW2008 in both programs. For a stop mass equal to the top mass the ratio of cross
sections is about 15\%, decreasing quickly as the stop mass increases. 

In fig.~\ref{fig:stop} (right) we show the ``double" ratio $R_{14/8}({\rm top+stop})/R_{14/8}({\rm top})$, where $R_{14/8}(X)$ is the ratio of the cross-section for producing final state $X$ at the LHC 14 and 8 TeV. Such cross-section ratios have been introduced~\cite{Mangano:2012mh} due to their very high theoretical precision (since most of the theoretical uncertainties cancel), 
and because they can be accurately measured.
Unfortunately, as can be seen from fig.~\ref{fig:stop} (right), this particular double ratio has size that is at most few permil, which likely makes it experimentally inaccessible. 

The reason for this double ratio's smallness is that top and stop production are both dominated by
$gg$ scattering and scale in a similar way with the center of mass energy, which is the result of two competing factors. First, as discussed in Ref.~\cite{Mangano:2012mh}, the BSM contribution can be accessed in such a ratio when the BSM signal and the SM background are dominated by different parton luminosities (which is not the case here). 
Second, the different masses of tops and stops lead to different scalings with the c.m. energy. This latter factor, alone, ensures that in the general case the cross section ratios have some sensitivity to BSM dynamics even if it is initiated by the same parton luminosity as the SM background.

\acknowledgments
We thank Michelangelo Mangano for many insightful discussions and helpful suggestions.  
The work of M.C. and P.F. was supported by the DFG Sonderforschungsbereich/Transregio 9 ÒComputergest\"utzte Theoretische TeilchenphysikÓ. M.C. was also supported by the Heisenberg programme of the Deutsche Forschungsgemeinschaft. The work of A.M. is supported by ERC grant 291377 ``LHCtheory: Theoretical predictions and analyses of LHC physics: advancing the precision frontier". J.R. is supported by a Marie Curie Intra--European Fellowship of the European Community's 7th Framework Programme under contract number PIEF-GA-2010-272515.

\end{document}